\documentclass[twocolumn,showpacs,preprintnumbers]{revtex4}
\usepackage{amsmath}
\usepackage{graphicx}
\usepackage{dcolumn}
\usepackage{bm}
\usepackage{subfigure}
\usepackage{times}
\begin{document}
\title{Decoherence-free creation of atom-atom  entanglement in cavity via  fractional adiabatic passage}
\author{Mahdi Amniat-Talab$^{1,2}$}
\email{amniyatm@u-bourgogne.fr}
\author{St\'{e}phane Gu\'{e}rin$^{1}$}
\email{sguerin@u-bourgogne.fr}
\author{Hans-Rudolf Jauslin$^{1}$}
\email{jauslin@u-bourgogne.fr}
 \affiliation{$^{1}$Laboratoire de Physique, UMR CNRS 5027,
Universit\'{e} de Bourgogne, B.P. 47870, F-21078 Dijon,
France.\\$^{2}$Physics Department, Faculty of Sciences, Urmia
University, P.B. 165, Urmia, Iran.}
\date{\today }
\begin{abstract}

We propose a  robust and decoherence insensitive scheme to
generate controllable  entangled states of two three-level atoms
interacting with an optical cavity and a laser beam. Losses  due
to atomic spontaneous transitions and to cavity decay are
efficiently suppressed by employing fractional adiabatic passage
and appropriately designed atom-field couplings. In this scheme
the two atoms traverse the cavity-mode and the laser beam in
opposite directions as opposed to other entanglement schemes in
which the atoms are required to have fixed locations inside a
cavity.  We also show that the coherence of a traveling atom can
be transferred to the other one without populating the
cavity-mode.
\end{abstract}
 \maketitle

The physics of entanglement provides the basis of applications
such as quantum information processing and quantum communications.
Particles can then be viewed as carriers of quantum bits of
information and the realization of engineered entanglement is an
essential ingredient of the implementation of quantum gates
\cite{qgate}, cryptography \cite{EkertPRL91} and teleportation
\cite{BennettPRL93}. The creation of long-lived entangled pairs of
atoms may provide reliable quantum information storage. The idea
is to apply a set of controlled coherent interactions to the atoms
of the system in order to bring them into a tailored entangled
state. The problem of controlling entanglement is thus directly
connected to the problem of coherent control of population
transfer in multilevel systems.


In the context of cavity QED, one of the main obstacles to
realize atom-atom entanglement  is the  decoherence resulting from
the cavity decay. Additionally, the cavity couples to an excited
state of the atom that undergoes spontaneous emission. Regarding
these considerations, in recent years several schemes to entangle
atoms \cite{PlenioPRA99,SorensenPRL03} and to implement quantum
gates
\cite{BeigePRL00,PachosPRL02,JanePRA02,YiPRL03,PellizzariPRL95}
using optical cavities have been proposed. To avoid decoherence
effects, it is most convenient to design transfer strategies that
do not populate the decoherence channels during the time-evolution
of the system.
\begin{figure}
\label{setup}
\includegraphics[width=6cm]{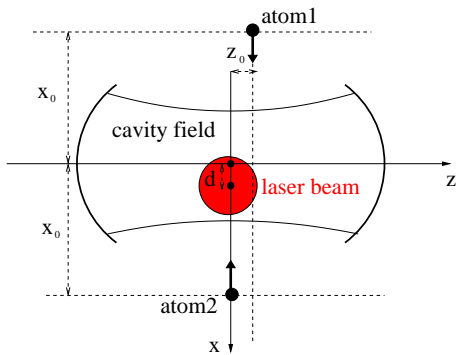}
\caption{Geometrical configuration  of the  atoms-cavity-laser
system in the proposed scheme. The propagation direction of the
laser beam is parallel to $y$ axis and  perpendicular to the
page.}
\end{figure}

In this paper we propose an alternative way to entangle two
traveling
 atoms interacting with an optical cavity and a laser
beam, based on 3-level interactions in a $\Lambda$-configuration.
This method is based on the coherent creation of superposition of
atom-atom-cavity states via fractional stimulated Raman adiabatic
passage (f-STIRAP) \cite{VitanovJPB99}, that keeps the cavity-mode
and the excited atomic states unpopulated during the whole
interaction. In Ref. \cite{AmniatPRA05}, using f-STIRAP and a
one-atom dark state,  a robust scheme to generate atom-atom
entanglement was proposed, where the two traveling atoms encounter
the cavity-mode one by one. Here, the two atoms enter
simultaneously into the cavity in such a way that the system
follows a two-atom dark state that allows us to keep additionally
the cavity-mode empty.

We consider the situation described in Fig. 1,  where the two
atoms move in planes orthogonal to the $z$ axis as follows:
\begin{eqnarray}\label{coords}
    z_{1}&=&z_{0},~~x_{1}=-x_{0}+v_{1}t\cos\theta_{1},~~y_{1}=-y_{0}+v_{1}t\sin
    \theta_{1},\nonumber\\
z_{2}&=&0,~~x_{2}=x_{0}+v_{2}(t-\tau)\cos\theta_{2},\nonumber\\
y_{2}&=&-y_{0}+v_{2}(t-\tau)\sin
    \theta_{2},
\end{eqnarray}
where $(x_{i},y_{i},z_{i}),i=1,2$ are the coordinates of the
$i$-th atom $(x_{0},y_{0}>0)$,  $\tau$ is the time-delay of the
second atom with respect to the first one, $\theta_{i}$ is the
angle that the $i$-th atom constructs with the positive direction
of the $x$ axis ($\theta_{1}\in [0,\pi/2[,~\theta_{2}\in
]\pi/2,\pi]$), and $v_{i}$ is the velocity of the $i$-th atom.
Since the laser field propagates in direction of the $y$ axis, the
time-dependent optical phase  of the laser field seen by each atom
is
\begin{equation}\label{opt-phases}
\varphi_{1}(t)=\omega_{L} t-kv_{1}t\sin \theta_{1},\quad
\varphi_{2}(t-\tau)=\omega_{L} t-kv_{2}(t-\tau)\sin \theta_{2},
\end{equation}
where $k$ is the wavevector magnitude of the laser field. Figure 2
represents the linkage pattern of the
atom-cavity-laser system. The laser pulse associated to the Rabi frequency $%
\Omega (t)$ couples the states $|g_{1}\rangle $ and $|e\rangle$,
and the
cavity-mode  with Rabi frequency $G(t)$ couples the states $%
|e\rangle$ and $|g_{2}\rangle$. The Rabi frequencies $\Omega(t)$
  and $G(t)$ are chosen real without loss
of generality. These two fields interact with the atom with a time
delay, each of the fields is in one-photon resonance with the
respective transition. The semiclassical Hamiltonian of this
system in the resonant approximation where
\begin{equation}\label{RWA}
|\Omega_{0}|, |G_{0}|\ll
\omega_{e},\omega_{C},|\partial\varphi_{1}/\partial
t|,|\partial\varphi_{2}/\partial t|,
\end{equation}
with $\Omega_{0},G_{0}$ the peak values of the Rabi frequencies,
can then be written  as ($\hbar=1$)
\begin{figure}\label{linkage}
\includegraphics[width=6cm]{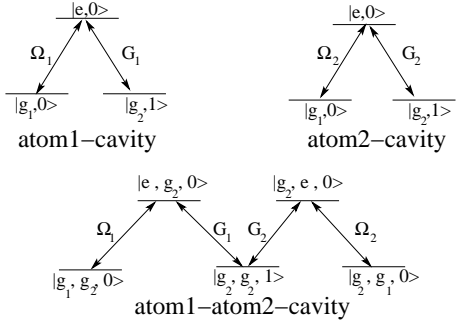}
\caption{Linkage pattern of  the  system corresponding to the
effective Hamiltonian. }
\end{figure}
\begin{eqnarray}\label{H}
    H(t)&=&\sum_{i=1,2}\big[\omega_{e}|e\rangle_{ii}\langle
    e|+\big(G_{i}(t)~a|e\rangle_{ii}\langle
    g_{2}|+\text{H.c.}\big)\nonumber\\
    &+&\big(\Omega_{i}(t)e^{i\varphi_{i}(t)}|g_{1}\rangle_{ii}\langle
    e|+\text{H.c.}\big)\big]+\omega _{C}a^{\dag}a,
\end{eqnarray}
where the subscript $i$ on the states denotes the two atoms, $a$
is the annihilation operator of the cavity mode, $\omega _{e}$ is
the energy of the atomic excited state $(\omega _{g_{1}}=\omega
_{g_{2}}=0)$, and $ \omega _{C}$ is the frequency of the cavity
mode taking resonant $\omega _{C}=\omega _{L}=\omega _{e}$ which
implies $kv_{i}\sin\theta_{i}\ll \omega_{L}$. In the following we
consider the state of the atom1-atom2-cavity system as
$|A1,A2,n\rangle$ where $\{A1,A2=g_{1},e,g_{2}\}$, and $\{n=0,1\}$
is the number of photons in the cavity-mode.

Regarding Figure 2, the subspace $\mathcal{S}$ generated by the
states
$\{|g_{1},g_{2},0\rangle,|e,g_{2},0\rangle$,$|g_{2},g_{2},1\rangle,|g_{2},e,0\rangle$,$
|g_{2},g_{1},0\rangle\}$ is decoupled under $H$ from the rest of
the Hilbert space of the system. If we consider the initial state
of the system as $|g_{1},g_{2},0\rangle$, the  Hamiltonian of the
system in the subspace $\mathcal{S}$ will be
\begin{widetext}
\begin{equation}
    H_{P}(t):=PHP=\left(
\begin{array}{ccccc}
0 & \Omega_{1}(t)e^{+i\varphi_{1}(t)} & 0 & 0 & 0 \\
\Omega_{1} (t)e^{-i\varphi_{1}(t)} & \omega_{e} & G_{1}(t)& 0 & 0 \\
0 & G_{1}(t) & \omega_{C} & G_{2}(t-\tau) & 0 \\
0 & 0 & G_{2}(t-\tau) & \omega_{e} & \Omega_{2}(t-\tau)e^{-i\varphi_{2}(t-\tau)} \\
0& 0 & 0  & \Omega_{2}(t-\tau)e^{+i\varphi_{2}(t-\tau)}& 0
\end{array}%
\right),
\end{equation}
where  $P$ is the projector on the subspace $\mathcal{S}$. The
effective Hamiltonian is thus given by
\begin{equation}\label{Heff}
    H^{\text{eff}}:=R^{\dagger}H_{P}R-iR^{\dagger}\frac{\partial R}{\partial
    t}=\left[
\begin{array}{ccccc}
0 & \Omega_{1}(t) & 0 & 0 & 0 \\
\Omega_{1} (t) & kv_{1}\sin\theta_{1} & G_{1}(t)& 0 & 0 \\
0 & G_{1}(t) & kv_{1}\sin\theta_{1} & G_{2}(t-\tau) & 0 \\
0 & 0 & G_{2}(t-\tau) & kv_{1}\sin\theta_{1} & \Omega_{2}(t-\tau) \\
0& 0 & 0  & \Omega_{2}(t-\tau)&
k(v_{1}\sin\theta_{1}-v_{2}\sin\theta_{2})
\end{array}%
\right],
\end{equation}
\end{widetext}
where the unitary transformation $R$ is
\begin{equation}
    R=\left(
\begin{array}{ccccc}
1 & 0 & 0 & 0 & 0 \\
0 & e^{-i\varphi_{1}(t)} & 0& 0 & 0 \\
0 & 0 & e^{-i\varphi_{1}(t)} & 0 & 0 \\
0 & 0 & 0 & e^{-i\varphi_{1}(t)} &  \\
0& 0 & 0  & 0&e^{-i[\varphi_{1}(t)-\varphi_{2}(t-\tau)]}
\end{array}
\right).
\end{equation}
The  dynamics of the system is governed by the Schr\"{o}dinger
 equation $i\frac{\partial }{\partial t}|\Phi (t)\rangle
=H^{\text{eff}}(t)|\Phi (t)\rangle $.

An essential condition for the STIRAP and f-STIRAP processes is
the four-photon resonance between the states
$|g_{1},g_{2},0\rangle$ and $|g_{2},g_{1},0\rangle$ which means:
\begin{equation}\label{two-res-cond}
    |\Delta :=k(v_{1}\sin\theta_{1}-v_{2}\sin\theta_{2})|\ll
    |\Omega_{0}|,|G_{0}|.
\end{equation}
 This condition can be achieved by control of the
velocity of atoms, and of the deflection angles of the atoms
$\theta_{i=1,2}$. Numerics shows that, in practice, the condition
(\ref{two-res-cond}) is satisfied for
$\Delta\lesssim\{\Omega_{0},G_{0}\}/100$. Assuming this condition
allows one to consider $\Delta\sim kv_{1}\sin\theta_{1}\sim0$ in
Eq. (\ref{two-res-cond}), if we additionally assume  $v_{1}\sim
v_{2}\equiv v$. In this case $\tau$ is the delay of arrival at the
center of the cavity $(x=y=0)$ of the second atom with respect to
the first one.

The system is taken to be initially in the state
$|g_{1},g_{2},0\rangle$,
\begin{equation}\label{Phi-in}
|\Phi (t_{i})\rangle =|g_{1},g_{2},0\rangle.
\end{equation}
The goal is to transform it at the end of interaction into an
atom-atom entangled state
\begin{eqnarray}\label{Phi-fin}
 |\Phi ( t_{f} )\rangle&=&\cos \vartheta |g_{1},g_{2},0\rangle+\sin \vartheta
 |g_{2},g_{1},0\rangle\nonumber\\
 &=&\left(\cos \vartheta |g_{1},g_{2}\rangle+\sin \vartheta
 |g_{2},g_{1}\rangle\right)|0\rangle,
\end{eqnarray}
where  $\vartheta $ is a constant mixing angle $(0\leq \vartheta
\leq \pi /2)$, and the cavity-mode state factorizes and is left in
the vacuum state. The qubits are stored in the two degenerate
ground states of the atoms. The decoherence due to atomic
spontaneous emission is produced if the  states
$\{|e,g_{2},0\rangle,|g_{2},e,0\rangle\}$ are populated, and the
cavity decay occurs if the  state $|g_{2},g_{2},1\rangle$ is
populated during the adiabatic evolution of the system. Therefore
we will design the Rabi frequencies
$\{\Omega_{1}(t),G_{1}(t),\Omega_{1}(t),G_{2}(t)\}$ in our scheme
such that these states are not populated during the dynamics.

One of the instantaneous eigenstates (the two-atom dark state) of
$H^{\text{eff}}(t) $ which corresponds to a zero eigenvalue  is
\cite{PellizzariPRL95}
\begin{eqnarray}\label{dark}
|D(t)\rangle&=&C\Big(G_{1}\Omega_{2}|g_{1},g_{2},0\rangle
-\Omega_{1}\Omega_{2}|g_{2},g_{2},1\rangle\nonumber\\
&+& G_{2}\Omega_{1}|g_{2},g_{1},0\rangle\Big),
\end{eqnarray}%
where $C$ is a normalization factor. The possibility of
decoherence-free generation of atom-atom entanglement arises from
the following behavior of the dark  state:
\begin{subequations}
\begin{eqnarray}\label{Da}
  &&\lim_{t\rightarrow t_{i}}\frac{\Omega_{1} (t)}{\Omega_{2}(t)}=0,\qquad |D(t_{i})\rangle\sim |g_{1},g_{2},0\rangle\\\label{Db}
  \nonumber\\&&\lim_{t\rightarrow t_{f} }\frac{\Omega _{1}(t)}{\Omega _{2}(t)}=\tan \vartheta,\nonumber\\
  && |D(t_{f})\rangle\sim \cos \vartheta |g_{1},g_{2},0\rangle
  +\sin \vartheta|g_{2},g_{1},0\rangle\\\label{Dc}
  \nonumber\\&&~t_{i}< t< t_{f},~G_{1}(t)\sim G_{2}(t)\gg\Omega_{1} (t),\Omega_{2}(t),\nonumber\\
  && |D(t)\rangle\sim\Omega_{2}
  (t)|g_{1},g_{2},0\rangle+\Omega_{1}(t)|g_{2},g_{1},0\rangle.
\end{eqnarray}
\end{subequations}
Equations (\ref{Da}) and (\ref{Db}) are known as  f-STIRAP
conditions \cite{VitanovJPB99,AmniatPRA05}, and the condition
(\ref{Dc}) guarantees the absence of population in the state
$|g_{2},g_{2},1\rangle$ during the time-evolution of the system
\cite{VitanovPRA98,Steane}. Equation (\ref{Db}) means  that the
Rabi frequencies fall off in a constant ratio, during the  time
interval where they are non-negligible. We remark that this
formulation opens up the possibility to implement f-STIRAP with
Gaussian pulses. The goal in the following is to show that such a
pulse sequence can be designed in a cavity by an appropriate
choice of the parameters.

In an optical cavity, the spatial variation of the atom-field
coupling for the maximum coupling TEM$_{00}$ mode, resonant with
the $|e\rangle \leftrightarrow |g_{2}\rangle $ atomic transition,
is given by
\begin{equation}
G(x,y,z)=G_{0}~e^{-(x^{2}+y^{2})/W_{C}^{2}}\cos \left( \frac{2\pi z}{\lambda }%
\right),  \label{RabiC}
\end{equation}
where $W_{L}$ is the waist of the cavity mode, and
$G_{0}=-\mu\sqrt{\omega_{C}/(2\epsilon_{0}V_{\text{mode}})}$ with
$\mu$ and $V_{\text{mode}}$  respectively the dipole moment of the
atomic transition and the effective volume of the cavity mode. The
spatial variation of the atom-laser coupling for the laser beam of
Fig. 1 is
\begin{equation}\label{RabiL}
    \Omega(x,z)=\Omega_{0}~e^{-(x^{2}+z^{2})/W_{L}^{2}},
\end{equation}
where $W_{L}$ is the waist of the laser beam, and
$\Omega_{0}=-\mu\mathcal{E}/2$ with $\mathcal{E}$ the  amplitude
of the laser field. Figure 1 shows a situation where the first
atom, initially in the state $|g_{1}\rangle $, goes with velocity
$v$ (on the $y=0$ plane at $z=z_{0}$ line) through an optical
cavity initially in the vacuum state $|0\rangle$ and then
encounters the laser beam, which is parallel to the $y$ axis
(orthogonal to the cavity axis and the trajectory of the atom).
The laser beam is resonant with the $|e\rangle \leftrightarrow
|g_{1}\rangle$ transition. The distance between the center of the
cavity and the laser axis is $d$. The second atom, synchronized
with the first one $\tau=0$, moves with the same velocity $v$ on
the $y=0$ plane at $z=0$ in the opposite direction with respect to
the first atom. The traveling atoms encounter the time-dependent
and delayed Rabi frequencies of the cavity-mode and the laser
fields as follows
\begin{subequations}
\begin{eqnarray}
G_{1}(t) &=&G_{0}~e^{-(vt)^{2}/W_{C}^{2}}\cos \left( \frac{2\pi z_{0}}{\lambda }%
\right) , \\
\Omega_{1}(t)
&=&\Omega_{0}~e^{-z_{0}^{2}/W_{L}^{2}}~e^{-(vt-d)^{2}/W_{L}^{2}},\\
G_{2}(t) &=&G_{0}~e^{-(vt)^{2}/W_{C}^{2}},\\
\Omega_{2}(t)&=&\Omega_{0}~e^{-(vt+d)^{2}/W_{L}^{2}}
\end{eqnarray}
\end{subequations}
 where the time origin is defined when the atoms
meet the center of the cavity at $x=y=0$. The appropriate values
of $z_{0}$ and $d$ that lead to the f-STIRAP process can be
extracted from a contour plot of the final population
$P_{|g_{1},g_{2},0\rangle }(t_{f}):=|\langle g_{1},g_{2},0|\Phi
(t_{f} )\rangle |^{2}$ as a function of $z_{0}$ and $d$ that we
calculated numerically (see Fig. 3). The white dot in Fig. 3
 shows values of $z_{0}$ and $d$ to obtain an f-STIRAP process with $%
\vartheta \simeq \pi /4$ (called half-STIRAP). It has been chosen
such that at the end of interaction $P_{|g_{1},g_{2},0\rangle
}(t_{f})\simeq P_{|g_{2},g_{1},0\rangle }(t_{f})\simeq 0.5$, and
the populations of the other states of the subspace $\mathcal{S}$
are zero.

\begin{figure}\label{contour}
\includegraphics[width=7cm]{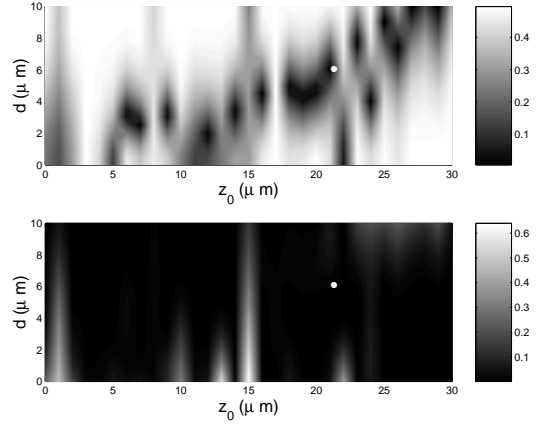}
\caption{Top panel: contour plot at the final time $t_{f}$ of $|\frac{1}{2}%
-P_{|g_{1},g_{2},0\rangle }(t_{f})|$ as a function of $z_{0}$ and
$d$ (black areas correspond to approximately half population
transfer) with the pulse
parameters as $W_{L}=20\protect\mu $m, $~W_{C}=40\protect\mu $m,$~v=2$ m/s,$~%
\protect\lambda =780$nm, $~\Omega _{0}=20\frac{v}{W_{L}},~G_{0}=100\frac{v}{%
W_{C}}$. Bottom panel: The same plot for the sum of the final
populations in intermediate states
$P_{|e,g_{2},0\rangle}(t_{f})+P_{|g_{2},g_{2},1\rangle}(t_{f})+P_{|g_{2},e,0\rangle}(t_{f})$
where black areas correspond to approximately zero population. The
white dot shows specific values of $z_{0}$ and $d$ used in Fig. 4
to obtain a half-STIRAP process.}
\end{figure}
\begin{figure}
\includegraphics[width=7cm]{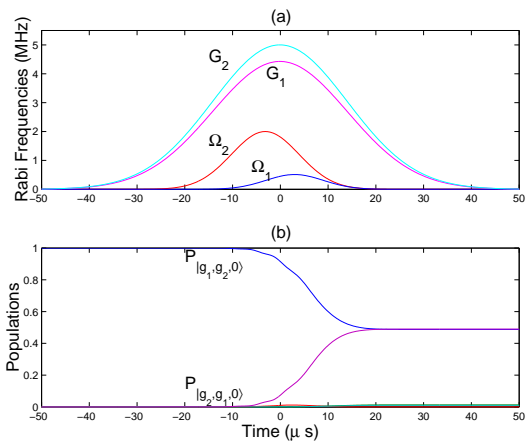}
\caption{(a) Rabi frequencies of the cavity-mode and the laser
field for two atoms.  (b) Time evolution of the populations  which
represents a two-atom half-STIRAP.} \label{Fstirap}
\end{figure}
\begin{figure}
\includegraphics[width=7cm]{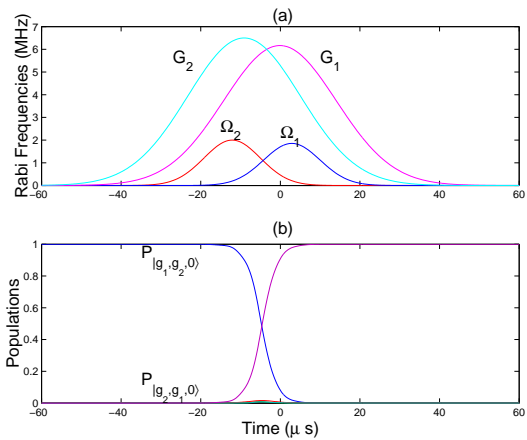}
\caption{(a) Rabi frequencies of the cavity-mode and the laser
field for two atoms with the pulse parameters as $v=2$ m/s,
$W_{L}=20\protect\mu$m, $W_{C}=40\protect\mu $m, $\lambda =780$
nm, $\Omega _{0}=2$ MHz, $G_{0}=6.5$ MHz, $z_{0}=5.5\protect\mu$m,
$d=6\protect\mu$m and $\tau=-9\protect\mu$s. (b) Time evolution of
the populations which represents a two-atom STIRAP.}
\label{Fstirap}
\end{figure}
Figure 4 shows for parameter values associated to the white dot in
Fig. 3,  (a) the time dependence of the Rabi frequencies of
half-STIRAP for two atoms, and (b) the time evolution of the
populations which shows that the population is split  at 50\%
among the states $|g_{1},g_{2},0\rangle ,|g_{2},g_{1},0\rangle $
and is almost zero for the other states of $\mathcal{S}$ during
the interaction for $G_{0}\sim3\Omega_{0}$. This case corresponds
to the generation of the maximally atom-atom entangled state
$1/\sqrt{2}\big(|g_{1},g_{2},0\rangle+|g_{2},g_{1},0\rangle\big)$
by adiabatic passage.
Assuming Gaussian pulse profiles for $\Omega (t)$ and $G(t)$ of widths $%
T_{L}=W_{L}/v$ and $T_{C}=W_{C}/v$ respectively,  the sufficient
condition of adiabaticity is $\Omega _{0}T_{L},G_{0}T_{C}\gg 1.$

The state $|g_{2},g_{2},0\rangle$ is stationary state of the
system. If the second atom enters inside the cavity before the
first one ($\tau<0$), we can transfer completely the population
from the state $|g_{1},g_{2},0\rangle$ to $|g_{2},g_{1},0\rangle$
\emph{without populating the other states of $\mathcal{S}$ during
the dynamics} (see Fig. 5). In particular, given that the first
atom initially is prepared in a coherent superposition of the
ground states $\alpha|g_{1}\rangle+\beta|g_{2}\rangle$, that the
second atom is initially in the state $|g_{2}\rangle$, and that
the cavity-mode is initially in the vacuum state, a two-atom
STIRAP will coherently map this superposition onto the second
atom:
\begin{equation}\label{map}
    \alpha|g_{1},g_{2},0\rangle+\beta|g_{2},g_{2},0\rangle\rightarrow\alpha|g_{2},g_{1},0\rangle+\beta|g_{2},g_{2},0\rangle.
\end{equation}

In summary, we have proposed a robust and decoherence-free scheme
to generate atom-atom entanglement, using the f-STIRAP technique
in $\Lambda$-systems. This scheme is robust with respect to
variations of the velocity of the atoms $v$, of the peak Rabi
frequencies $G_{0},\Omega_{0}$ and  of the field detunings, but
not with respect to the parameters $d,z_{0}$, describing the
relative positions of the laser beam and the cavity, shown in Fig.
1. Our scheme can be implemented in an optical cavity with
$G_{0}\sim\kappa\sim\Gamma$. The necessary condition to suppress
the cavity decoherence is $G_{0}\gg\Omega_{0}$ which is satisfied
in practice for $G_{0}\sim3\Omega_{0}$ (see Figs. 4,5). For given
values of $W_{C},W_{L}$, the adapted values of $d$ and $z_{0}$ in
the f-STIRAP process can be determined from a contour plot of the
final populations. Decoherence channels are suppressed during the
whole evolution of the system. In this scheme, as opposed to the
scheme of Ref. \cite{PellizzariPRL95}, we do not need to fix the
atoms inside the optical cavity nor to apply two laser beams for
each of the individual atoms.

M. A-T. wishes to acknowledge the financial support of the MSRT of
Iran and SFERE. We acknowledge support from the Conseil
R\'{e}gional de Bourgogne.
\bibliography{paper4}
\end{document}